\documentclass[prl,showpacs]{revtex4-2}
\usepackage{amsmath,amssymb}

\newcommand{\ee}{\mathrm{e}}
\newcommand{\ii}{\mathrm{i}}
\newcommand{\dd}{\mathrm{d}}

\begin{document}

\title{Comment on ``Landau Levels and Riemann Zeros''}

\author{Mi-Ra Hwang}
\affiliation{Department of Electronic Engineering, Kyungnam University, Changwon, 631-701, Korea}
\author{Eylee Jung}
\affiliation{Department of Electronic Engineering, Kyungnam University, Changwon, 631-701, Korea}
\author{MuSeong Kim}
\affiliation{Pharos iBio Co., Ltd. Head Office: \#1408, 38, Heungan-daero 427beon-gil, Dongan-gu, Anyang, 14059, Korea}
\author{DaeKil Park}
\email{dkpark@kyungnam.ac.kr}
\affiliation{Department of Electronic Engineering, Kyungnam University, Changwon, 631-701, Korea}
\affiliation{Department of Physics, Kyungnam University, Changwon, 631-701, Korea}

\date{\today}

\begin{abstract}
Sierra and Townsend (ST) [1] showed that the ``absorption spectrum'' model of Connes [2] arises as the lowest-Landau-level (LLL) limit of a charged particle in crossed electric and magnetic fields, and suggested that the higher Landau levels supply the fluctuation part of the Riemann-zero counting function $N(E)$. Within the approximation used by ST we construct the eigenfunctions of an arbitrary Landau level $n$ and derive their asymptotics at the two box boundaries that ST identify. The boundary condition is the same as for $n=0$ up to an $E$-independent phase $(-\ii)^n$, so no $\ln E$ term arises for any fixed $n$. We also note that the $\ln E$ term in the semiclassical Eq.~(29) of ST is the first-order shift of the smooth LLL counting function, $E\to E+\gamma z^2$.
\end{abstract}

\maketitle

\section{Setting}

We use the notation of ST with $\hbar=\ell=1$ and $\omega_c\gg|\omega_h|$. In this limit $H\simeq H_c+H_h$ with $[H_c,H_h]=0$, so the eigenstates are products of a Landau-level state $|n\rangle$ (energy $\gamma(n+\tfrac12)$, $\gamma=\omega_c/|\omega_h|$) and an eigenstate of $H_h=\tfrac12(QP+PQ)$ with eigenvalue $E$ (in units of $\hbar|\omega_h|$). For $n=0$, ST Eq.~(15) gives
\begin{equation}
\psi^{+}_{E,0}(x,y)=C\int_{-\infty}^{\infty}\!\dd Q\;\ee^{-\ii Qy}\,\ee^{-(x-Q)^2/2}\,\Phi^{+}_E(Q),
\qquad \Phi^{+}_E(Q)=|Q|^{-1/2+\ii E},
\label{eq:psi0int}
\end{equation}
which evaluates to
\begin{equation}
\psi^{+}_{E,0}(x,y)=C\,\ee^{-x^2/2}\,M(a,b,u),\qquad
a=\tfrac14+\tfrac{\ii E}{2},\quad b=\tfrac12,\quad u=\tfrac12(x-\ii y)^2 ,
\label{eq:psi0}
\end{equation}
with $M$ the confluent hypergeometric function [ST, Eq.~(16)]. The corresponding Landau-level energy is $\gamma(n+\tfrac12)$, where $\gamma=\omega_c/|\omega_h|$. The particle is confined to the box $|x|,|y|\le L$, and ST impose a condition relating $\psi(x,L)$ to $\psi(L,x)$. We treat the $+$ sector; the argument below uses only the form of the large-$|u|$ asymptotics of $M$ and therefore carries over to the $-$ sector with its own $(a,b)$.

\section{Eigenfunctions of the $n$-th Landau level}

The $n$-th Landau function centred at $Q$ is $H_n(x-Q)\,\ee^{-(x-Q)^2/2}$, with $H_n (z)$ the  Hermite polynomial. The natural generalisation of Eq.~\eqref{eq:psi0int} is
\begin{equation}
\psi^{+}_{E,n}(x,y)=C\int\!\dd Q\;\ee^{-\ii Qy}\,H_n(x-Q)\,\ee^{-(x-Q)^2/2}\,\Phi^{+}_E(Q).
\label{eq:psin}
\end{equation}

\paragraph{Operator form.} $H_n(x-Q)$ is a polynomial in $Q$, and $Q^k\ee^{-\ii Qy}=(\ii\partial_y)^k\ee^{-\ii Qy}$. Since the remaining factors in Eq.~\eqref{eq:psin} do not depend on $y$, and $x$ and $\partial_y$ commute,
\begin{equation}
\psi^{+}_{E,n}(x,y)=H_n\!\left(x-\ii\partial_y\right)\psi^{+}_{E,0}(x,y).
\label{eq:opform}
\end{equation}
(Note that this is \emph{not} $(x-\ii\partial_y)^n\psi_{E,0}$; the polynomial is $H_n$ of the operator, which contains the correct lower-order terms.) For instance $H_2(x-\ii\partial_y)=4x^2-8\ii x\partial_y-4\partial_y^2-2$.

\section{Asymptotics at the two boundary points}

For large $|u|$ [3]
\begin{align}
M(a,b,u)&\simeq \frac{\Gamma(b)}{\Gamma(b-a)}(-u)^{-a}\,[1+O(u^{-1})], && \mathrm{Re}\,u<0,\label{eq:Mneg}\\
M(a,b,u)&\simeq \frac{\Gamma(b)}{\Gamma(a)}\,\ee^{u}u^{a-b}\,[1+O(u^{-1})], && \mathrm{Re}\,u>0.\label{eq:Mpos}
\end{align}
Below $L\to\infty$ at fixed $n$ and fixed $x$; ``$\simeq$'' means up to relative corrections $O(1/L)$.

\paragraph{Point $(x,L)$.} Here $u=\tfrac12(x-\ii L)^2$, so $-u=\tfrac12(L+\ii x)^2$ and $\mathrm{Re}\,u<0$. Using Eq.~\eqref{eq:Mneg},
$\psi_{E,0}(X,Y)\simeq C\,\ee^{-X^2/2}\,\frac{\Gamma(1/2)}{\Gamma(1/4-\ii E/2)}\big[\tfrac12(Y+\ii X)^2\big]^{-a}$. Each $\partial_Y$ acts on the power law and gives $\partial_Y=-2a/(Y+\ii X)=O(1/L)$, while the prefactor $\ee^{-X^2/2}$ is independent of $Y$. Hence in Eq.~\eqref{eq:opform} only the $Y$-independent part of the operator survives:
\begin{equation}
\psi^{+}_{E,n}(x,L)\simeq C\,\ee^{-x^2/2}\,\frac{\Gamma(1/2)}{\Gamma(\tfrac14-\tfrac{\ii E}{2})}\left(\frac{L^2}{2}\right)^{-\frac14-\frac{\ii E}{2}}H_n(x).
\label{eq:xL}
\end{equation}

\paragraph{Point $(L,x)$.} Here $u=\tfrac12(L-\ii x)^2$ and $\mathrm{Re}\,u>0$. From Eq.~\eqref{eq:Mpos}, with $X\to L$, $Y\to x$ and using $-X^2/2+u=-Y^2/2-\ii XY$,
\begin{equation}
\psi_{E,0}(X,Y)\simeq C\,\frac{\Gamma(1/2)}{\Gamma(a)}\;\ee^{\varphi}F(X,Y),\qquad \varphi=-\tfrac12Y^2-\ii XY,\quad F=u^{a-1/2}.
\end{equation}
Now $\partial_Y\varphi=-Y-\ii X$ is $O(L)$, so the derivatives in $H_n(X-\ii\partial_Y)$ must \emph{not} be dropped on $\ee^{\varphi}$. Conjugating,
\begin{equation}
\ee^{-\varphi}\,(X-\ii\partial_Y)\,\ee^{\varphi}=\ii\,(Y-\partial_Y),
\end{equation}
and since conjugation is an algebra homomorphism,
\begin{equation}
\ee^{-\varphi}H_n(X-\ii\partial_Y)\ee^{\varphi}=H_n\big(\ii(Y-\partial_Y)\big).
\end{equation}
The $O(L)$ terms have thereby cancelled exactly. The function $F$ varies slowly, $\partial_Y\ln F=O(1/L)$, so to leading order $F$ may be treated as a constant, and we need $H_n(\ii(Y-\partial_Y))\cdot 1$. Because $Y$ and $\partial_Y$ do not commute, this is not $H_n(\ii Y)$. Use the generating function $\sum_n \frac{t^n}{n!}H_n(w)=\ee^{2wt-t^2}$ with $w=\ii(Y-\partial_Y)$. With $A=2\ii tY$, $B=-2\ii t\partial_Y$ and $[A,B]=-4t^2$, the Baker--Campbell--Hausdorff formula gives
\begin{equation}
\ee^{A+B-t^2}\,1=\ee^{A}\ee^{B}\ee^{-[A,B]/2-t^2}\,1=\ee^{2\ii tY+t^2}
=\sum_n\frac{t^n}{n!}\,\ii^{\,n}H_n(Y),
\end{equation}
where we used $\ee^{B}1=1$. Therefore
\begin{equation}
H_n\big(\ii(Y-\partial_Y)\big)\,1=\ii^{\,n}H_n(Y),
\end{equation}
and
\begin{equation}
\psi^{+}_{E,n}(L,x)\simeq C\,\ii^{\,n}\,\ee^{-x^2/2-\ii xL}\,\frac{\Gamma(1/2)}{\Gamma(\tfrac14+\tfrac{\ii E}{2})}\left(\frac{L^2}{2}\right)^{-\frac14+\frac{\ii E}{2}}H_n(x).
\label{eq:Lx}
\end{equation}
(The corrections from $\partial_Y F$ are of relative order $n/L$.)

\section{Boundary condition}

Both Eq.~\eqref{eq:xL} and Eq.~\eqref{eq:Lx} are proportional to $H_n(x)$, which cancels in the ratio (the statement is uniform in $x$ by analyticity in $x$; at the zeros of $H_n$ both sides vanish at this order and the identity follows by continuity):
\begin{equation}
\frac{\psi^{+}_{E,n}(x,L)}{\psi^{+}_{E,n}(L,x)}
=(-\ii)^n\,\ee^{\ii xL}\,\frac{\Gamma(\tfrac14+\tfrac{\ii E}{2})}{\Gamma(\tfrac14-\tfrac{\ii E}{2})}\left(\frac{L^2}{2}\right)^{-\ii E}.
\label{eq:ratio}
\end{equation}
For $n=0$ this is the ratio used by ST. For general $n$ the $E$-dependent part is unchanged; the only new ingredient is the constant $(-\ii)^n$. The identification of $(x,L)$ with $(L,x)$ in ST carries an $E$-independent phase $\ee^{\ii\theta_n}$ (its value depends on how the identification is extended to $n>0$), so the quantisation condition becomes
\begin{equation}
2\theta(E)-E\ln\frac{L^2}{2\pi}=\theta_n+\frac{n\pi}{2}\pmod{2\pi},
\end{equation}
with $\theta(E)=\arg\Gamma(\tfrac14+\tfrac{\ii E}{2})-\tfrac E2\ln\pi$ the Riemann--Siegel function. The left-hand side is identical for all $n$. Consequently the counting function for each Landau-level tower differs from the $n=0$ one at most by an $E$-independent constant, and the full spectrum is the $n=0$ spectrum shifted by $\gamma(n+\tfrac12)$. No $\ln E$ term, and no fluctuation, is generated by the higher Landau levels at fixed $n$.

\section{Remark on Eq.~(29) of ST}

Let $f(E)=\frac{E}{2\pi}\ln\frac{L^2}{2\pi}-\frac{E}{2\pi}\ln\frac{E}{2\pi}+\frac{E}{2\pi}$ be the smooth LLL counting function. Then
\begin{equation}
f'(E)=\frac{1}{2\pi}\Big[\ln\frac{L^2}{2\pi}-\ln\frac{E}{2\pi}\Big],
\end{equation}
and the terms of Eq.~(29) proportional to $\gamma z^2$ are exactly $\gamma z^2f'(E)$. Thus Eq.~(29) equals $f(E+\gamma z^2)$ to first order: the semiclassical treatment shifts the energy argument by the cyclotron energy $\gamma z^2$. This is a smooth, deterministic term with a fixed sign for large $E$, whereas the fluctuation $S(E)=\tfrac1\pi\arg\zeta(\tfrac12+\ii E)$ changes sign and is only bounded in modulus by $O(\ln E)$. The $\ln E$ term of Eq.~(29) therefore does not have the character of a fluctuation.

\section{Conclusion and scope}

Within the $\omega_c\gg|\omega_h|$ approximation of ST, the exact eigenfunctions of every fixed Landau level $n$ satisfy the same boundary condition as the LLL, up to the $E$-independent phase $(-\ii)^n$. The $\ln E$ term of Eq.~(29) is not supported by this quantum calculation. Our derivation holds for fixed $n$ as $L\to\infty$; the regime in which $n$ grows with $L$ or $\gamma$, corresponding to large classical amplitude $z$, is not covered by it and remains open. Beyond the approximation $H\simeq H_c+H_h$ the exact model is still a two-degree-of-freedom quadratic (hence integrable) Hamiltonian, which has no exponentially proliferating family of periodic orbits of lengths $\ln p$ of the type required for a fluctuation term in the sense of Berry.

\begin{acknowledgments}
D.P.\ thanks Claude (Anthropic) for assistance with the calculations in Secs.~2--4.
\end{acknowledgments}

\end{document}